\documentclass[sigconf, nonacm]{acmart}

\usepackage{float}
\usepackage{xcolor}
\usepackage{tcolorbox}
\usepackage{comment}
\usepackage{rotating}
\usepackage[super]{nth}
\usepackage{multirow} 
\usepackage{colortbl}
\usepackage{nicematrix}
\usepackage{subcaption,booktabs}
\usepackage{float}
\usepackage[flushleft]{threeparttable}
\usepackage{algpseudocode}
\usepackage{makecell}

\usepackage{pbox}
\usepackage{diagbox}
\usepackage{subcaption}
\usepackage[linesnumbered,ruled,vlined]{algorithm2e}
\usepackage{footnote}
\makesavenoteenv{algorithm}
\usepackage{amsthm}
\usepackage{url}
\usepackage{booktabs}
\usepackage{makecell}

\definecolor{my_blue_color}{RGB}{137,168,245}

\algrenewcommand\algorithmicrequire{\textbf{Input:}}
\algrenewcommand\algorithmicensure{\textbf{Output:}}
\newcolumntype{L}[1]{>{\raggedright\arraybackslash}m{#1}}
\newcolumntype{C}[1]{>{\centering\arraybackslash}m{#1}}
\newcolumntype{R}[1]{>{\raggedleft\arraybackslash}m{#1}}
\newcolumntype{P}[1]{>{\centering\arraybackslash}p{#1}}
\newcommand{\smallfootnote}{\fontsize{8.1}{11}\selectfont}

\newtheorem{definition}{Definition}
\newtheorem{lemma}{Lemma}

\newcommand\vldbyear{2026}
\newcommand\vldbworkshop{Tabular Data Analysis (TaDA)}
\newcommand\vldbauthors{\authors}
\newcommand\vldbtitle{\shorttitle} 
\newcommand\vldbavailabilityurl{https://github.com/soroushomidvar/LDI}
\newcommand\vldbpagestyle{empty} 

\begin{document}
\title{LDI: Localized Data Imputation for Text-Rich Tables}

\author{Soroush Omidvartehrani}
\email{s.omidvartehrani@ualberta.ca}
\affiliation{%
  \institution{University of Alberta}
  \city{Edmonton}
  \country{Canada}
}

\author{Davood Rafiei}
\email{drafiei@ualberta.ca}
\affiliation{%
  \institution{University of Alberta}
  \city{Edmonton}
  \country{Canada}
}







\begin{abstract}
Missing values are pervasive in real-world tabular data and can significantly impair downstream analysis. Imputing them is especially challenging in text-rich tables, where dependencies are implicit, complex, and dispersed across long textual fields. Recent work has explored using Large Language Models (LLMs) for data imputation, yet existing approaches typically process entire tables or loosely related contexts, which can compromise accuracy, scalability, and explainability. We introduce LDI, a novel framework that leverages LLMs through localized reasoning, selecting a compact, contextually relevant subset of attributes and tuples for each missing value. This targeted selection reduces noise, improves scalability, and provides transparent attribution 
by revealing the dependency relations that justify each selected attribute and the evidence behind each retrieved tuple. It makes clear not only which data influenced a prediction, but also why it was chosen.
Through extensive experiments on real and synthetic datasets, we demonstrate that LDI consistently outperforms state-of-the-art imputation methods, achieving up to 8\% higher accuracy with hosted LLMs and even greater gains with small local models. The improved interpretability and robustness also make LDI well-suited for high-stakes data management applications.
\end{abstract}

\maketitle

\pagestyle{\vldbpagestyle}
\begingroup\small\noindent\raggedright\textbf{VLDB Workshop Reference Format:}\\
\vldbauthors. \vldbtitle. VLDB \vldbyear\ Workshop: \vldbworkshop.\\ 
\endgroup
\begingroup
\renewcommand\thefootnote{}\footnote{\noindent
This work is licensed under the Creative Commons BY-NC-ND 4.0 International License. Visit \url{https://creativecommons.org/licenses/by-nc-nd/4.0/} to view a copy of this license. For any use beyond those covered by this license, obtain permission by emailing \href{mailto:info@vldb.org}{info@vldb.org}. Copyright is held by the owner/author(s). Publication rights licensed to the VLDB Endowment. \\
\raggedright Proceedings of the VLDB Endowment. 
ISSN 2150-8097. \\
}\addtocounter{footnote}{-1}\endgroup

\vspace{-3mm}

\ifdefempty{\vldbavailabilityurl}{}{
\vspace{.3cm}
\begingroup\small\noindent\raggedright\textbf{VLDB Workshop Artifact Availability:}\\
The source code, data, and/or other artifacts have been made available at \url{https://github.com/soroushomidvar/LDI}.
\endgroup
}

\section{Introduction}

Missing values are a common characteristic of many real-world datasets, particularly those derived from open or heterogeneous sources. If left unaddressed, they can significantly undermine the reliability of downstream analysis and machine learning applications. For example, many learning algorithms operate under the assumption that all input data is fully observed~\cite{emmanuel2021survey,zhang2008missing, kumar2017data}, and their performance often degrades when faced with incomplete records. This not only reduces predictive accuracy but can also introduce systematic bias into analytical outcomes~\cite{kang2013prevention, zhang2007gbkii, qin2007semi, stoyanovich2020responsible}. As a result, predicting or imputing missing values is a critical component of the broader data wrangling process, often helping to preserve the underlying structure of the data and ensuring the validity of subsequent analysis~\cite{song2020imputing}.

\smallskip
\noindent\textbf{Motivation.}
While the problem of data imputation has been studied for decades, a growing fraction of modern datasets are text-rich tables that mix structured fields with lengthy textual descriptions, reviews, or metadata. These tables are common in domains such as e-commerce, real estate, scientific data management, and open government data. Unlike purely numerical datasets, text-rich tables contain dependencies that are implicit, complex, and often buried within long, noisy text. Identifying which textual cues are informative for imputation remains a major open challenge.

\begin{figure*}[tb!]
  \centering
  \includegraphics[width=\linewidth]{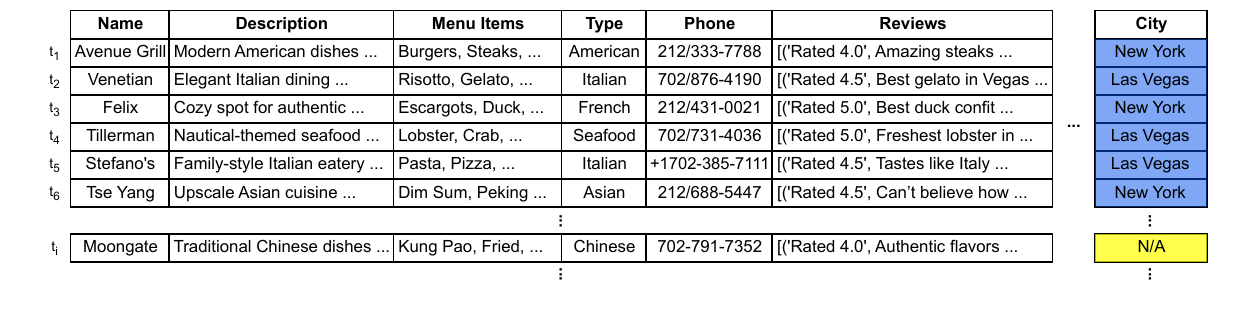}
  \caption{Example of data imputation}
  \label{fig:ex}
\end{figure*}

\smallskip
\noindent\textbf{Example.} 
Consider a restaurant table with missing values in the \texttt{City} column (see Figure~\ref{fig:ex}). Several other attributes---including \texttt{Description}, \texttt{Phone}, and \texttt{Reviews}---might help infer the correct city, but their relevance varies.
The \texttt{Description} column may contain location-related hints (e.g., ``Modern American dishes'' and ``Elegant Italian dining'') that are typically too general or vague to be useful, while the \texttt{Reviews} field may offer more specific but noisy references (e.g., ``Best gelato in Vegas''). 
The \texttt{Phone} column, in contrast, encodes reliable geographic signals via area codes, though inconsistencies in formatting can obscure them. Determining which attributes and records truly contribute to accurate imputation requires both contextual understanding and selective reasoning.

\smallskip
\noindent\textbf{Limitations of Existing Work.}
Existing approaches to data imputation range from simple heuristics to advanced generative models. Heuristic techniques (e.g., mean or mode imputation) are fast but ignore inter-attribute dependencies, often leading to biased  estimates~\cite{baraldi2010introduction,khan2020sice}.
Model-based methods, including regression, ex\-pec\-ta\-tion-\-max\-i\-mi\-za\-tion (EM), and matrix factorization, capture statistical dependencies but can be computationally expensive and rely on strong assumptions about the data distribution~\cite{little2019statistical}. 
Recent deep learning–based approaches, such as autoencoders, GANs, and transformer models, have shown promise in capturing nonlinear patterns, especially in large or multi-modal datasets~\cite{yoon2018gain}. Building on this progress, large language models (LLMs) have emerged as powerful tools for data imputation, thanks to their ability to model complex dependencies and generalize across diverse contexts~\cite{narayan2022can, zhang2024jellyfish, li2024towards}.
However, these approaches remain fundamentally global in scope as they tend to process entire tables or large portions thereof. This introduces three major issues that remain largely overlooked: 
(1)~\textit{Scalability}: Tables often exceed model context windows, forcing either truncation or splitting that loses important structure.
(2)~\textit{Noise sensitivity}: Including loosely related rows and columns dilutes the signal, making imputations less accurate. 
(3)~\textit{Lack of transparency}: When the entire table is fed to an LLM, it becomes unclear which attributes or tuples influenced each prediction, undermining interpretability and user trust.

These limitations raise an important but underexplored question: \emph{Can missing values be accurately imputed using only a localized subset of the table---one that preserves essential context while improving scalability and explainability?}

\smallskip
\noindent\textbf{Our Approach.}
We hypothesize that, in many cases, missing values can be imputed accurately using a localized context---a compact subset of attributes and tuples most relevant to the missing value. Our approach, LDI, operationalizes this idea by decomposing the problem into two sub-tasks: (1)~selecting a subset of columns that are most relevant to the column with missing values, and (2)~identifying a subset of rows that provide sufficient contextual evidence for imputation.
This localized reasoning enables LDI to reduce noise, scale to larger tables, and trace which data influenced each prediction.
Our extensive evaluation across multiple datasets demonstrates that LDI not only improves performance but also enhances interpretability and robustness, particularly in text-rich, noisy, and heterogeneous environments. 

\smallskip
\noindent\textbf{Contributions:}
Our contributions can be summarized as follows:
(1)~We introduce LDI, a framework that performs attribute- and tuple-level selection to guide LLM-based imputation.
(2)~By limiting context to relevant subsets, LDI achieves both higher accuracy and transparent attribution.
(3)~LDI handles textual inconsistencies and heterogeneous formats more effectively than existing models.
(4)~The framework supports small, local LLMs, enabling efficient and privacy-preserving imputation.
(5)~Through analytical and experimental evaluations on real-world datasets from various domains, we demonstrate that LDI achieves superior accuracy, scalability, and explainability compared to state-of-the-art methods.

\section{Related Work}\label{sec:related_works}

Our work relates to two lines of research: data imputation, which targets the recovery of missing values, and dependency detection, which isolates the columns most predictive of a given target.

\subsection{Data Imputation}

Missing data imputation has been extensively studied, leading to a wide range of approaches that vary in complexity, assumptions, and applicability.

\paragraph{Traditional Methods.}
Early imputation techniques replace missing values using fixed constants, such as zero or the mean, or heuristic strategies like Last Observation Carried Forward (LOCF) and Next Observation Carried Backward (NOCB) \cite{engels2003imputation,lachin2016fallacies,klamroth2014three}. While computationally inexpensive, these methods rely on strong assumptions about missingness patterns and data ordering, often resulting in biased imputations in real-world datasets \cite{molnar2008does}.

\paragraph{Statistical Models.}
More advanced approaches estimate the underlying data distribution to predict missing values. Representative methods include relational dependency networks \cite{mayfield2010eracer}, factor-graph-based models \cite{rekatsinas2017holoclean,wu2020attention}, distance likelihood maximization \cite{song2020imputing}, and low-rank matrix completion \cite{candes2012exact}. Non-linear models, such as tree-based ensembles
\cite{stekhoven2012missforest,chen2016xgboost}, further improve flexibility. However, these techniques often assume specific data distributions or linear relationships, limiting their robustness in heterogeneous or noisy datasets.

\paragraph{Similarity-based Approaches.}
Similarity-based methods infer missing values using neighboring tuples with complete information. Classic techniques include k-Nearest Neighbors (KNN) \cite{altman1992introduction} and ensemble variants such as KNNE \cite{domeniconi2004nearest}. Other approaches, such as MIBOS \cite{wu2012missing}, rely on iterative similarity counting, while clustering-based methods perform imputation within clusters
formed using similarity measures \cite{zhang2008missing,yan2015missing,nikfalazar2017new}. Although effective in some settings, these methods depend heavily on predefined similarity functions and struggle with semantic similarity and free-text data.

\paragraph{Generative Models.}
Recent advances in deep generative models have enabled more expressive imputation techniques. GAIN \cite{yoon2018gain} applies adversarial training, while MIDA \cite{gondara2018mida} and HI-VAE \cite{nazabal2020handling} leverage autoencoders and variational inference.
More recently, language-model-based approaches such as IPM \cite{mei2021capturing}, LakeFill \cite{yang2025data}, ZeroEC \cite{wu2025zero}, UnIMP \cite{wang2025llm}, and FMW \cite{narayan2022can} exploit semantic representations to improve imputation quality. Despite their strong performance, these methods generally operate as black boxes and provide limited insight into why a particular value was imputed.

\subsection{Dependency Detection}
Dependency discovery has long been studied in the context of relational data. Functional dependencies (FDs) capture exact deterministic relationships between attributes~\cite{papenbrock2015functional}, while approximate functional dependencies (AFDs) relax this requirement to tolerate noise \cite{kivinen1995approximate,mandros2017discovering}.
FDs and AFDs form the basis of many data cleaning and repair techniques, including conditional FDs \cite{bohannon2006conditional}, scalable FD discovery \cite{papenbrock2016hybrid}, and regression-based dependency learning \cite{guo2019learning}. They have also been applied to guide imputation and candidate selection \cite{king2003discovery,breve2022renuver,song2020enriching, shraga2023explaining, amsterdamer2023selecting}.
However, these approaches are primarily designed for structured attributes and assume consistent formatting and exact or near-exact matches.
They are less effective for free-text attributes, partial matches, and semantically related values. In contrast, LDI leverages transformation-based pattern discovery and LLM-based reasoning to capture dependencies that are not expressible using traditional FDs or AFDs.

\section{Preliminary}

We formalize the problem of missing data imputation and outline the setting for LDI.

\smallskip
\noindent \textbf{Problem Statement.}
Let $\mathcal{D}$ be a dataset with schema $\mathcal{R}$ consisting of attributes $A_j \in \mathcal{R}$. Each tuple $t_i \in \mathcal{D}$ contains a set of cell values, where a missing entry in the $j$-th attribute is denoted by $t_i[A_j] =\varnothing$.  
Let $\hat{\mathcal{D}}$ denote the imputed version of $\mathcal{D}$ such that, for all $i, j$ where $t_i[A_j]$ is missing in $\mathcal{D}$, the corresponding imputed tuple $\hat{t_i} \in \hat{\mathcal{D}}$ satisfies $\hat{t_i} [A_j] \neq \varnothing$. We denote by $\mathcal{D}^*$ the ground truth dataset in which each tuple $t_i^*\in\mathcal{D}^*$ contains the true values $t_i^*[A_j]$ for all attributes $A_j\in\mathcal{R}$.

\smallskip
\noindent \textbf{Goal.}
Given a dataset $\mathcal{D}$, the objective is to produce an imputed dataset $\hat{\mathcal{D}}$ by predicting missing entries of a target attribute $A_T$, reconstructing the ground truth dataset $\mathcal{D}^*$. LDI focuses on identifying the most relevant tuples and attributes for each imputation task and leveraging the reasoning capabilities of LLMs to achieve the following: (1) high imputation accuracy in real-world datasets, even when missing values depend on complex and unknown relationships; (2) improved interpretability by tracing which parts of the input influenced each prediction; (3) robustness to noise and formatting inconsistencies without relying on strict assumptions; and (4) minimal dependence on external data or task-specific resources, enabling broader and more efficient applicability.



\smallskip
\noindent \textbf{Assumptions.}
We assume that (1)~some values of $A_T$ (the attribute with missing entries) appear more than once across tuples, allowing the model to learn associations from these repetitions, (2)~one or more textual or categorical attributes are correlated with $A_T$ through textual, structural, or contextual cues, and (3) the observed (non-missing) values are sufficiently informative to support imputation.
Similar assumptions are commonly made in prior work on data imputation~\cite{rekatsinas2017holoclean, breve2022renuver, rezig2021horizon, zhang2008missing}.

\section{Proposed Method}
\begin{figure*}[bt!]
  \centering
  \includegraphics[width=\linewidth]{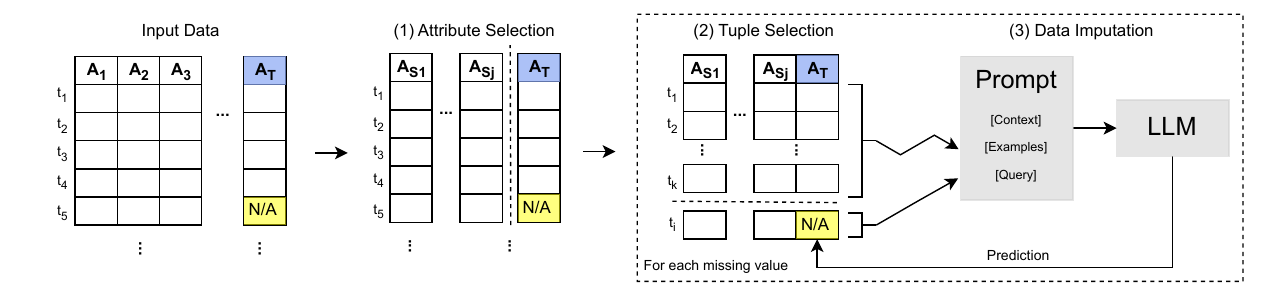}
  \caption{Overview of the LDI framework}
  \label{fig:main}
\end{figure*}

While LLMs' strong performance on a range of table reasoning tasks makes them promising candidates for data imputation, they often struggle with scalability to large tables~\cite{chen2022large, ji2023survey} and lack interpretability in their imputation decisions.
To address these issues, LDI, selectively identifies the most relevant attributes and tuples, focusing the model’s attention and improving explainability. 
In the first phase ($\S$\ref{sec:relevant_attribute_identification}), LDI retains only the attributes that are likely related to the target attribute $A_T$ and informative for imputation. The second phase ($\S$\ref{sec:similar_tuple_selection}) filters tuples to produce a relevant yet diverse set of records. Finally, in the third phase ($\S$\ref{sec:imputation-using-llm}), the selected attributes and tuples are fed into the LLM, which performs the imputation. An overview of this process is demonstrated in Figure~\ref{fig:main}. 

\subsection{Attribute Selection} \label{sec:relevant_attribute_identification}

To identify attributes most relevant to the imputation target, LDI performs attribute selection in two main phases: (1) detecting group-specific patterns in candidate columns, and (2) evaluating approximate dependency relationships. 
While traditional functional dependencies can model relationships between attributes, they are often too strict for real-world tables that contain noise and long textual values. To better handle such cases, we introduce a relaxed dependency criterion tailored to noisy, text-rich data.

\begin{definition}[\textbf{(p, q)-Approximate Dependency}]\mbox{}\\
Let \( X \) and \( A \) be columns in a table, and let \( p, q \in [0,1] \).
We say that a $(p, q)$-approximate dependency $X \rightarrow A$ holds if, for at least a fraction $p$ of the distinct values $\alpha$ in column $A$, there exists a common substring $s$ in column $X$ such that:
\begin{itemize}
 \item The probability that $s$ appears in a record given the value $\alpha \in A$ is at least $q$, i.e., $P(s \in X \mid A = \alpha) \geq q$, and
 \item For all $	\alpha' \neq \alpha$, $P(s \in X \mid A = 	\alpha') < q$.
\end{itemize}
\end{definition}

This relaxed definition captures partial and group-specific relationships between attributes. It relies on substring-level evidence,  making it robust to formatting inconsistencies and local variations. The parameters \( p \) and \( q \) allow the strictness of the dependency to be tuned, enabling a balance between robustness and sensitivity to weak signals.
We use this criterion to identify a subset of candidate attributes \( A_C \) that exhibit a $(p, q)$-approximate dependency with the target attribute \( A_T \) with the caveat that such dependencies, like traditional functional dependencies, cannot be definitively inferred from a database instance, and one can only verify that a possible dependency is not violated.

\subsubsection{Pattern-Based Dependency Detection} 

To assess whether $A_T$ is dependent on a candidate attribute $A_C$, LDI searches for group-specific patterns in $A_C$. Tuples are grouped by distinct values of $A_T$, and within each group we extract substrings from $A_C$ using the Longest Common Substring (LCS).
Intuitively, the LCS of the values in a group surfaces the substring they share, which serves as a candidate group-specific pattern (e.g., the common area-code prefix of phone numbers within a single city).
Substrings appearing in at least a $q$ fraction of tuples in a group are retained.
To satisfy dependency constraints, a retained substring must be unique to a single group; substrings frequent in multiple groups are discarded. We also remove redundant substrings that are fully contained in longer ones, simplifying interpretation without affecting correctness.
For example, phone numbers associated with Las Vegas frequently share prefixes such as \mbox{‘702/’}, while New York numbers share \mbox{‘212/’} (see \texttt{Phone} column in Figure~\ref{fig:ex}). After filtering cross-group and redundant substrings, only distinctive patterns (e.g., \mbox{‘702/’} and \mbox{‘212/’}) remain.
The LCS method is particularly effective in uncovering non-trivial patterns, even in the presence of noises such as typos or formatting inconsistencies. Its robustness stems from two key properties:
(1)~If a pattern is missing in some cells of a group, the LCS across the remaining cells is preserved. Adjusting the parameter $q$ makes it possible to relax the matching requirement while maintaining the detection of meaningful dependencies.
(2)~Inconsistent formatting produces shorter but still relevant LCS results. For example, variations such as \mbox{‘+1780 ’}, \mbox{‘780/’}, and \mbox{‘780-’}, still reveal the underlying pattern \mbox{‘780’}.

\subsubsection{Dependency Evaluation}

LDI next evaluates whether a $(p,q)$-approximate dependency holds between $A_C$ and $A_T$. A dependency is accepted if at least a fraction $p$ of groups contain at least one unique pattern---i.e., a substring that frequently appears within that group but is uncommon across others. This ensures dependencies are supported by consistent group-level evidence rather than isolated or random occurrences. 
For instance, when predicting \texttt{City} from \texttt{Phone}, if unique patterns exist for four out of five cities, the dependency holds for any $p \le 0.8$. Attributes satisfying this criterion are retained for subsequent phases.




\subsubsection{Scalability through Stratified Group Sampling}
Although our method can efficiently detect dependencies on large datasets (see $\S$\ref{sec:ex-runtime} and \ref{app:attr}), those with hundreds of thousands or millions of records may still require additional measures to ensure scalability. To this end, we introduce an optional \emph{group sampling} step.
While LDI can operate on a random sample of tuples, naive random sampling is inadequate as it overrepresents frequent values in $A_T$ and underrepresents rare ones, which can obscure group-level patterns. 
Since LDI relies on identifying such patterns, it is essential to preserve the diversity of $A_T$ in the sample. 

\paragraph{Group Sampling.}
To maintain diversity in the target attribute $A_T$, we partition the table by distinct values of $A_T$ and then sample in a structured way. We randomly select $m$ groups that each contain at least $n$ rows, and within each group, we randomly draw $n$ rows, producing a sample of $m \times n$ tuples.
When the dataset lacks sufficient tuples for a full $m \times n$ sample, such as when some values of $A_T$ are infrequent, we include as many complete groups as possible, prioritizing those with larger sizes. Specifically, we first select all groups with at least $n$ rows, then add those with $n-1$ rows, continuing until no further groups can be added without exceeding the sample limit.
The parameters $m$ (number of groups) and $n$ (tuples per group) control the breadth and depth of the sample. Increasing $m$ covers more distinct $A_T$ values, while increasing $n$ captures finer-grained or noisier within-group patterns.
Smaller values of $(m, n)$ may suffice for more homogeneous datasets. 
This stratified approach preserves group-level diversity while keeping the sample size manageable, enabling LDI to scale effectively without sacrificing the reliability of dependency detection.

\subsection{Tuple Selection} \label{sec:similar_tuple_selection}
In the second phase, a small set of example tuples is selected to guide the LLM during imputation. The goal is to choose tuples that provide strong contextual evidence while remaining representative and avoiding bias. 
Specifically, selected tuples must:
(1) have complete values for the target attribute $A_T$,
(2) be similar to the tuple with the missing value, and
(3) cover diverse values of $A_T$ to reduce bias.

We first retain tuples with non-missing values in $A_T$, which are required as valid examples\footnote{Tuples may have missing values in other attributes since full completeness is not required, but they are less likely to be similar to the selected tuple for imputation.}.
Similarity is computed only over the relevant attributes identified in the previous phase, avoiding noise from irrelevant fields.
For a tuple $t_i$ with missing $A_T$, similarity to a complete tuple $t_j$ is measured using the average normalized Longest Common Substring (LCS) over the selected attributes $S$:


\begin{equation}
\text{similarity}(t_i, t_j) = \frac{1}{|S|} \sum_{A \in S} \frac{|\text{LCS}(t_i[A], t_j[A])|}{\max(|t_i[A]|, |t_j[A]|)}
\label{eq:avg_lcs_similarity}
\end{equation}

Unlike embedding-based similarity, which can miss fine-grained patterns in long text, LCS captures exact overlapping substrings and preserves key lexical structure, making it effective for identifying closely related tuples.
Tuples are ranked by similarity, and the top $k$ tuples with distinct $A_T$ values are selected. This balances relevance with diversity, ensuring that the LLM receives $k$ informative yet non-redundant retrieved examples for imputation.


Suppose the relevant attribute selected in the first phase for our running example is \texttt{Phone}, and we want to impute the missing \texttt{City} for tuple $t_i$ corresponding to \texttt{Moongate} in Figure~\ref{fig:ex}. 
\noindent We compute the pairwise similarity of $t_i$ with each complete tuple based on their \texttt{Phone} values as follows:

\vspace{2mm}

{\smallfootnote 
\begin{center}
\begin{tabular}{|
>{\columncolor{my_blue_color}}C{8mm}|C{12mm}|C{12mm}|C{15mm}|C{12mm}|}
\hline
\rowcolor{my_blue_color} 
Pair & LCS & Length & Max Length & Similarity \\
\hline
$t_i$, $t_1$ & \texttt{-7}     & 2 & 12 & 0.167 \\
$t_i$, $t_2$ & \texttt{702}    & 3 & 12 & 0.250 \\
$t_i$, $t_3$ & \texttt{02}     & 2 & 12 & 0.167 \\
$t_i$, $t_4$ & \texttt{702}    & 3 & 12 & 0.250 \\
$t_i$, $t_5$ & \texttt{702-}   & 4 & 14 & 0.286 \\
$t_i$, $t_6$ & \texttt{7}      & 1 & 12 & 0.083 \\
\hline
\end{tabular}
\end{center}
}

\vspace{2mm}

\noindent Ranking by similarity yields
$t_5$ (0.286; \texttt{City} = Las Vegas), followed by $t_2$ and $t_4$ (0.250; \texttt{City} = Las Vegas), then $t_1$ and $t_3$ (0.167; \texttt{City} = New York), and finally $t_6$ (0.083; \texttt{City} = New York).
When \( k = 1 \), the most similar tuple \( t_5 \) is selected.  
When \( k = 2 \), the algorithm chooses \( t_5 \) and \( t_1 \) to preserve similarity while covering different \texttt{City} values.
Although $t_2$ and $t_4$ are more similar to $t_i$ than $t_1$, they share the same \texttt{City} value as $t_5$ (Las Vegas); the diversity constraint therefore skips them in favor of $t_1$, the most similar tuple with a different \texttt{City} value (New York).

\subsection{Data Imputation Using LLM}
\label{sec:imputation-using-llm}
In the third phase of LDI, we leverage LLMs to impute missing values in a retrieval-augmented setting. The model is provided with a small set of example tuples that include the relevant attributes identified in \S~\ref{sec:relevant_attribute_identification} and retrieved similar tuples from \S~\ref{sec:similar_tuple_selection}.
The LLM then infers the missing value by recognizing patterns across these examples. 
Following prior work~\cite{narayan2022can}, each tuple is serialized as a sequence of key--value pairs. The prompt includes a brief task description, example tuples with observed target values, and a query tuple with a missing value. The full prompt template is provided in Appendix~\ref{app:prompt}.
LLMs are well suited to this final step as they leverage pretraining knowledge to ground predictions in real-world facts without requiring external lookup tables (e.g., associating area codes with cities), generalize to patterns not seen in the provided examples, and remain robust to noise and formatting variants (such as ‘+1-702’, ‘702/’, and ‘702-’).

\begin{figure*}[bt]
  \centering
  \includegraphics[width=\linewidth]{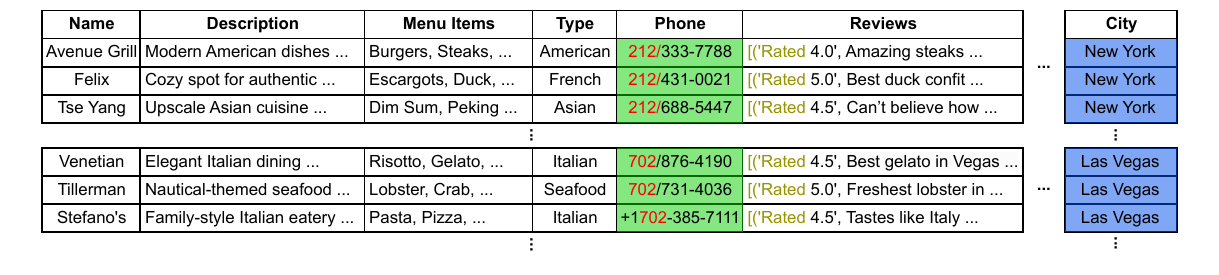}
  \caption{Example of data imputation (after applying our approach)}
  \label{fig:ex_solv}
\end{figure*}

\vspace{2mm}
\noindent\textbf{Running Example Recap.}
Suppose we aim to impute the missing \texttt{City} value for the tuple $t_i$ corresponding to \texttt{Moongate} in Figure~\ref{fig:ex}. The method first groups the data by \texttt{City} and samples $m$ groups with $n$ tuples each (Figure~\ref{fig:ex_solv}). Pattern detection identifies recurring substrings in candidate columns. Here, both \texttt{Phone} and \texttt{Reviews} show repeated substrings, but only \texttt{Phone} satisfies the group-level dependency criteria: area codes are unique to cities, whereas patterns in \texttt{Reviews} (e.g., ``\textit{[('Rated }'') appear across multiple cities and are discarded.
Next, similarity search identifies $k$ tuples most relevant to $t_i$ based on the selected attribute (\texttt{Phone}) while ensuring diversity in \texttt{City} values. These tuples are serialized into key-value format and provided to the LLM as retrieved context for retrieval-augmented imputation. The model then imputes the missing city by relying solely on informative signals in the \texttt{Phone} column, producing predictions that are both accurate and traceable.
This workflow demonstrates how LDI reduces noise, focuses on informative data, and leverages LLM reasoning to produce explainable imputations, even without predefined dependencies in the dataset.

\section{Performance Analysis}\label{sec:performance}

The runtime of LDI depends on the size of the dataset $|\mathcal{D}|$, sampled groups $m$, tuples sampled per group $n$, candidate attributes $a_c$, selected attributes $a_s$, complete tuples $c$, and average string length $\ell$. The following discussion breaks down the runtime of each phase. Details and proofs are provided in Appendix~\ref{app:analysis}.

\paragraph{Attribute Selection.}
For each candidate attribute, LDI extracts group-specific frequent patterns across the sampled groups using generalized suffix trees.
This phase runs in $\mathcal{O}(|\mathcal{D}| + a_c \cdot m \cdot \ell \cdot (n + r))$, where $r$
denotes the extracted patterns per group.
With redundancy filtering enabled, the complexity becomes
$\mathcal{O}(|\mathcal{D}| + a_c \cdot m \cdot \ell \cdot (n + r^2))$.

\paragraph{Tuple Selection.}
For each incomplete tuple, LDI computes similarity scores against the $c$ complete tuples using GST-based longest common substrings over the $a_s$ selected attributes and then selects the top-$k$ diverse examples. This results in a cost of $\mathcal{O}(a_s \cdot c \cdot \ell + c \log c)$ time per incomplete tuple.

\paragraph{LLM-based Imputation.}
LDI makes one LLM call per missing value. Each prompt contains a short fixed-length context, $k$ examples, and a query over $a_s$ attributes (see Figure~\ref{fig:main}). The token cost per prompt is $\lambda + k \cdot a_s \cdot \tau$, where $\lambda$ is the context length and $\tau$ is the average token length per attribute.

\section{Experiments}\label{sec:exp}
This section presents a comprehensive evaluation of our proposed data imputation framework, LDI. The experiments aim to answer three main questions:
(1) How does attribute and tuple selection affect imputation accuracy and explainability?
(2) How does LDI compare to existing state-of-the-art baselines?
(3) How robust and scalable is LDI across datasets and missingness conditions?

\subsection{Datasets and Experimental Setup}
Table~\ref{tab:dataset} summarizes the datasets. They vary in size, number of attributes, and missingness characteristics, and are widely used in data imputation studies~\cite{narayan2022can, chen2024seed, song2020imputing, mei2021capturing}, providing a diverse testbed. 
Experiments were conducted on a standard machine with 64 GB RAM and a 12-core CPU. Results are averaged over five runs. LLM-based imputation primarily used GPT-4o-mini, with Llama 3.2 3B used to evaluate performance in limited-resource settings. All remaining details are provided in Appendix~\ref{app:exp-setup}.


\begin{table}[tb]
\caption{Dataset summary}
\label{tab:dataset}
\smallfootnote
\begin{tabular}{|
>{\columncolor{my_blue_color}}C{12mm}|C{18mm}|C{9mm}|C{7mm}|C{21mm}|}
\hline
\rowcolor{my_blue_color} 
Dataset & Domain & \#Rows & \#Attr. & Incomplete Attr. (Vocabulary Size)\\ \hline
Buy & Digital Retail & 651 & 4 & manufacturer (62) \\
Restaurant & Food Services & 864 & 5 & city (49) \\
Zomato & Food Services & 51717 & 17 & location (93) \\
Phone & Mobile Devices & 413840 & 6 & brand name (385) \\ \hline
\end{tabular}
\end{table}

\subsection{Data Reduction via Attribute Selection}
Table~\ref{tab:data_reduction} highlights the efficiency of our model in reducing data size by focusing only on relevant attributes. Unlike other methods that use all available attributes, such as FMW, our approach selects a smaller set of dependent attributes identified in the initial phase. This targeted selection leads to a significant reduction in the amount of data processed, as shown by the lower token counts across all datasets. 
The reduction in tokens is computed by comparing the average number of tokens of the selected attributes to that of all attributes. The effect is substantial across all benchmarks: in the \textit{Zomato} dataset, token input is reduced by 96.1\% through the selection of only 2 out of 16 attributes. Similarly, the \textit{Buy}, \textit{Restaurant}, and \textit{Phone} achieve reductions of 63.8\%, 32.7\%, and 75.5\%, respectively. 
As a result, the process becomes more explainable and the outcomes more traceable, since it is easier to understand which attributes contribute to the final results.


\begin{table}[]
\smallfootnote
\centering
\caption{Average reduction in number of attributes and tokens}
\label{tab:data_reduction}
\begin{tabular}{|
>{\columncolor{my_blue_color}}C{13mm}|C{22mm}C{22mm}|C{12mm}|}
\hline
\cellcolor{my_blue_color} & \multicolumn{2}{c|}{\cellcolor{my_blue_color}\#Attr. (Except Target)} & \cellcolor{my_blue_color} \\ \cline{2-3}
\multirow{-2}{*}{\cellcolor{my_blue_color}Dataset} & \multicolumn{1}{c|}{\cellcolor{my_blue_color}All Attr. (FMW)} & \cellcolor{my_blue_color}Dep. Attr. (LDI) & \multirow{-2}{*}{\cellcolor{my_blue_color}\makecell{Less Data\\(\#Tokens)}} \\ \hline  
Buy & \multicolumn{1}{c|}{3} & 1 & 63.8\% \\
Restaurant & \multicolumn{1}{c|}{4} & 2 & 32.7\% \\
Zomato & \multicolumn{1}{c|}{16} & 2 & 96.1\% \\
Phone & \multicolumn{1}{c|}{5} & 1 & 75.5\% \\ \hline
\end{tabular}
\end{table}

\subsection{Performance Compared to Baselines}

We compare LDI with both traditional and LLM-based baselines: KNNE~\cite{domeniconi2004nearest} and MIBOS~\cite{wu2012missing} (nearest-neighbor-based), CMI~\cite{zhang2008missing} (clustering-based), ERACER~\cite{mayfield2010eracer}, Baran~\cite{mahdavi2020baran}, and HoloClean~\cite{rekatsinas2017holoclean} (statistics-based), along with two approaches that use language models: IPM~\cite{mei2021capturing} and FMW~\cite{narayan2022can}.
Baran, originally an error correction model, is adapted for data imputation by replacing each missing value with a null token and predicting its correction. The number of labels (examples) is set to 20, consistent with prior studies~\cite{mahdavi2020baran, mei2021capturing}.
To simulate missing data, we apply a \textit{missing completely at random} mechanism to originally complete datasets, removing 10\% of the values.
Table~\ref{tab:main-res} reports the accuracy of LDI compared to these baselines across four datasets.
Our method achieves the highest accuracy on three datasets (\textit{Buy}, \textit{Restaurant}, and \textit{Phone}) and the second-highest on \textit{Zomato}. LDI, IPM, and FMW all utilize language models and demonstrate high performance across all datasets. This creates a clear gap between them and traditional methods, which may perform reasonably on a single dataset but fail to generalize. For instance, Baran and HoloClean perform well only on \textit{Zomato}, with substantially lower results on the other datasets.

\begin{table*}[tb]
\smallfootnote
\centering
\caption{LDI accuracy compared to baselines at 10\% missing rate}
\label{tab:main-res}
\begin{tabular}{|
>{\columncolor{my_blue_color}}C{12.5mm}|C{13.5mm}|C{13.5mm}|C{13.5mm}|C{13.5mm}|C{13.5mm}|C{13.5mm}|C{14mm}|C{15mm}|C{14mm}|}
\hline
\rowcolor{my_blue_color} 
Dataset & KNNE & MIBOS & CMI & ERACER & Baran & HoloClean & IPM & FMW~(k=10) & LDI~(k=10) \\ \hline
Buy & 0.355±0.016 & 0.016±0.011 & 0.653±0.029 & 0.003±0.006 & 0.227±0.173 & 0.162±0.043 & \cellcolor{yellow!50}0.965±0.015 & 0.928±0.033 & \cellcolor{green!50}\textbf{0.974±0.026} \\
Restaurant & 0.206±0.019 & 0.093±0.025 & 0.560±0.049 & 0.185±0.031 & 0.315±0.044 & 0.331±0.056 & \cellcolor{yellow!50}0.772±0.023 & 0.766±0.122 & \cellcolor{green!50}\textbf{0.832±0.054} \\
Zomato & 0.581±0.005 & 0.024±0.001 & 0.746±0.006 & 0.001±0.000 & 0.938±0.023 & 0.956±0.005 & \cellcolor{green!50}\textbf{0.995±0.001} & 0.946±0.028 & \cellcolor{yellow!50}0.974±0.011 \\
Phone & 0.466±0.000 & 0.000±0.000 & 0.404±0.000 & 0.175±0.000 & 0.162±0.073 & 0.130±0.055 & 0.867±0.002 & \cellcolor{yellow!50}0.944±0.031 & \cellcolor{green!50}\textbf{0.970±0.016} \\ \hline
\end{tabular}
\end{table*}

IPM treats imputation as a classification task and introduces two variants: IPM-Multi and IPM-Binary\footnote{Under the IPM column, we report the better result of these two for each dataset.}. IPM-Multi formulates the problem as multiclass classification over the candidate set, and IPM-Binary ranks each candidate using a binary classification model. Both require fine-tuning to capture the relation between context and candidate values, making the model resource-intensive and less flexible. Moreover, IPM performance is sensitive to the quality of training data and the chosen modeling strategy. In practice, large and inconsistent gaps are often observed between IPM-Multi and IPM-Binary across datasets. In contrast, LDI avoids the need for training data by dynamically selecting informative attributes and constructing retrieval-augmented prompts using relevant examples. This allows LDI to adapt to new datasets more easily and generate more explainable localized prompts that are better aligned with the LLM's strengths.

FMW performs imputation by prompting a language model with tuples from the dataset, where examples can be randomly or manually selected. All attributes are included in the prompt, regardless of their relevance, which can introduce noise and distract the model. While FMW avoids training, is simple to apply, and allows manual selection of informative examples, this selection does not scale to large datasets\footnote{We therefore use random selection for comparison.} and their method does not distinguish between informative and irrelevant attributes. In contrast, LDI narrows the context to relevant tuples and attributes only, reducing prompt noise and enabling imputations that are both more accurate and easier to explain through clearly identifiable supporting signals.

\subsection{Robustness to Different Missingness Levels} \label{sec:missing-rate}
LDI is designed to maintain high accuracy regardless of the missing value rate, as long as it is provided with a few well-chosen examples. 
For best results, these examples should be sufficient to detect dependencies and also provide contextual information to frame the imputation task. This behavior is similar to Programming by Examples (PBE) methods, where the model generalizes well from a small, informative set of examples.

While the imputation task is defined over a single target column, additional missing values may exist in other attributes for both the query tuple and the candidate pool. Given that similarity is evaluated on a per-column basis, the presence of missing values diminishes the set of attributes contributing to the similarity computation. This reduction leads to lower similarity scores for such tuples, decreasing their chance of being selected during example retrieval.

\begin{figure}[bt!]
  \centering
  \includegraphics[width=\linewidth]{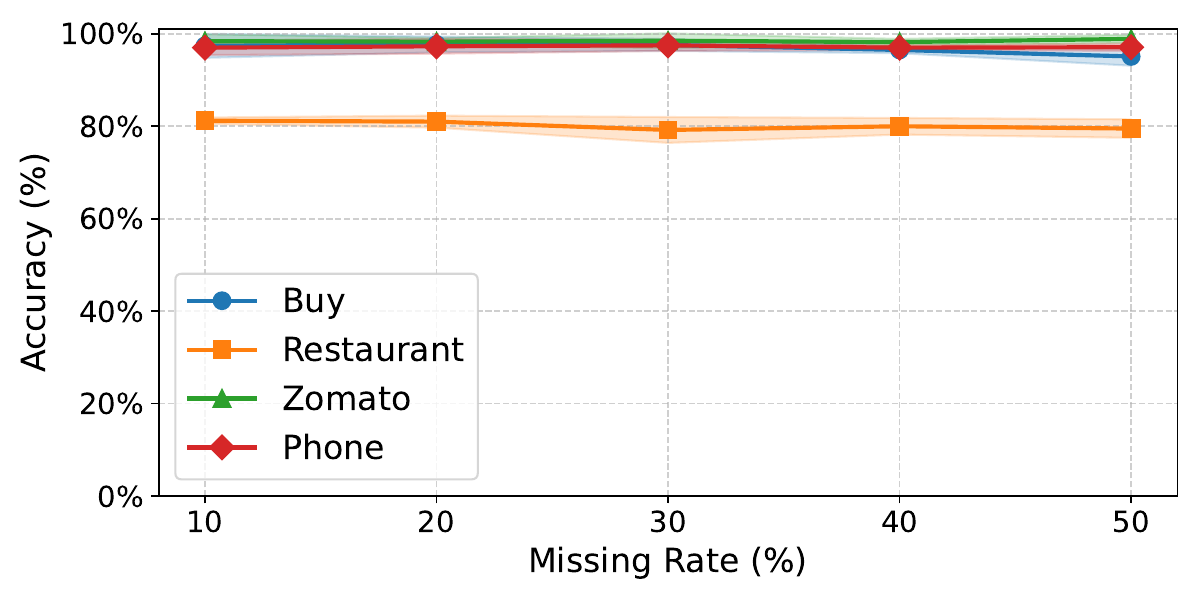}
  \caption{Impact of varying missing rates on accuracy}
  \label{fig:rates}
\end{figure}

To evaluate the robustness of LDI to changes in missing rate, we vary the missing rate from 10\% to 50\%, with a \textit{missing completely at random} mechanism applied to complete tuples. As shown in
Figure~\ref{fig:rates}, even after imputing 50\% of the data, the accuracy does not drop significantly, indicating that the model can still identify attribute dependencies and use similar examples to guide the LLM.
This robustness is expected to extend to different missingness mechanisms---\textit{missing at random} and \textit{missing not at random}---because the model can uncover attribute dependencies even from limited samples, as reflected in the results, and a small set of informative examples is sufficient to drive accurate imputations.

\subsection{Runtime and Scalability} \label{sec:ex-runtime}
Our results show that LDI is highly efficient and scalable. The attribute selection phase completes in negligible time across all datasets, and the tuple selection phase scales linearly with the number of imputations. Runtime grows moderately with increasing tuple counts, string lengths, and number of selected attributes, but remains practical even for thousands of tuples and long strings. 

Table~\ref{tab:runtime} summarizes execution times on real-world datasets. Attribute selection, executed once to identify dependent columns,  is consistently fast with slightly higher times for \textit{Zomato} and \textit{Phone} due to their larger number of attributes and longer average string lengths. Tuple selection must be performed for each missing value, and its runtime depends mainly on the number of tuples compared during similarity search as well as the number and length of selected attributes.

\begin{figure}[tb]
    \centering
    \begin{subfigure}[b]{0.49\columnwidth}
        \centering
        \includegraphics[width=\textwidth]{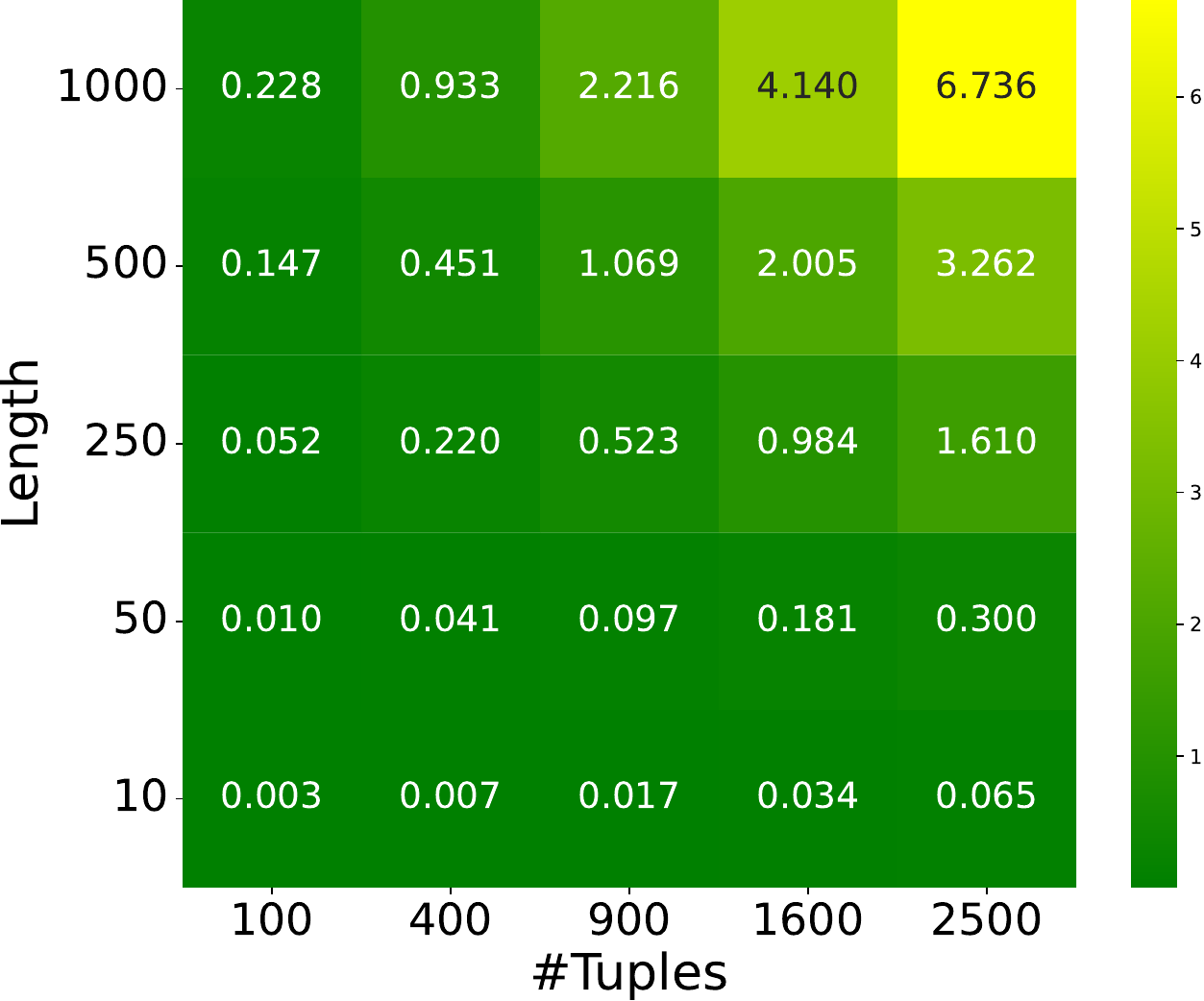}
        \caption{Attr. Selection}
        \label{fig:p1-runtime}
    \end{subfigure}
    \hfill
    \begin{subfigure}[b]{0.49\columnwidth}
        \centering
        \includegraphics[width=\textwidth]{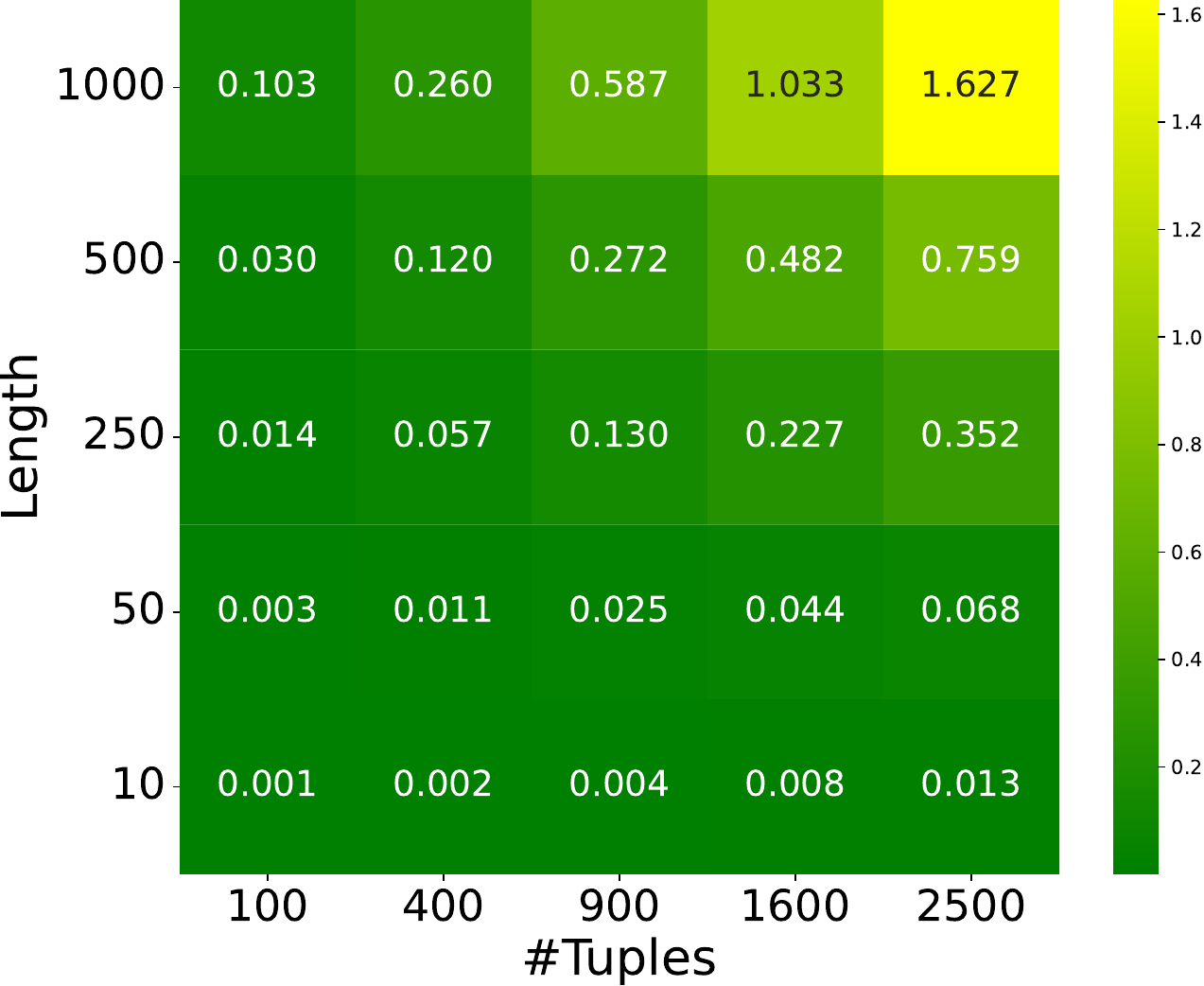}
        \caption{Tup. Selection}
        \label{fig:p2-runtime}
    \end{subfigure}
    \caption{Runtime trends on synthetic data (in seconds)}
    \label{fig:runtime}
\end{figure}

To further assess scalability with respect to input characteristics---the number of groups ($m$), tuples per group ($n$), and attribute value length in characters ($l$)---we conducted controlled experiments on synthetic data. We varied $l$ from 10 to 1000 and both $m$ and $n$ from 10 to 50. As shown in Section~\ref{sec:performance}, these parameters have the largest impact on the runtime of attribute selection. The number of returned substrings was controlled by setting $p=q=0.5$. The results are shown in Figure~\ref{fig:runtime}. As expected, runtime increases with both the number of tuples and the average string length during the attribute selection phase (Figure~\ref{fig:p1-runtime}), but remains practical even for lengthy strings and thousands of tuples. Group sampling can further bound this cost by operating on a small yet representative subset of the data; for instance, sampling only 100 tuples ($m=n=10$) was sufficient in our experiments to detect reliable patterns. The runtime of the tuple selection phase exhibits a similar trend (Figure~\ref{fig:p2-runtime}), growing moderately as the number and length of tuples increase. Since this phase must be repeated for every missing entry, its cost scales linearly with the number of imputations. When multiple attributes are selected in the first phase, the runtime increases proportionally, as similarity is computed pairwise across attributes. Nevertheless, even in these settings, runtime remains well within practical limits, supporting the scalability of our approach.

\begin{table}[tb]
\smallfootnote
\centering
\caption{Runtime on real-world datasets (in seconds)}
\begin{tabular}{|
>{\columncolor{my_blue_color}}C{14mm}|C{20mm}|C{20mm}|}
\hline
\rowcolor{my_blue_color} 
Dataset & Attr. Selection & Tup. Selection \\ \hline
Buy        & 0.022 & 0.058 \\ \hline
Restaurant & 0.012 & 0.116 \\ \hline
Zomato     & 0.338 & 0.375 \\ \hline
Phone      & 0.795 & 0.162 \\ \hline
\end{tabular}
\label{tab:runtime}
\end{table}

\subsection{Ablation Study}
To analyze the contribution of each phase, we evaluate performance by first using all attributes and randomly selected tuples, then adding attribute and tuple selection step by step. This layered setup shows how each phase improves results. Unlike typical ablation studies that remove one component at a time, we do not remove the first phase entirely, since the second phase depends on its selected attributes, and evaluating similarity over all attributes would obscure the effect of relevant signals due to noisy and redundant fields.
We evaluate performance using two metrics: exact match accuracy and \mbox{ROUGE-1~F1} score. The former measures whether the model prediction is exactly correct, while the latter provides a softer measure of correctness, crediting partial matches. Typically, ROUGE is used to evaluate summarization and generation tasks, and \mbox{ROUGE-1}, specifically, calculates the overlap of unigrams between the prediction and the reference. In our case, using \mbox{ROUGE-1~F1} helps distinguish between predictions that are completely incorrect and those that contain part of the correct value or provide a more verbose version of it. We report \mbox{ROUGE-1} instead of higher-order \mbox{ROUGE-N} metrics, as most target values are short (1-2 tokens), making unigram overlap a more reliable measure.

\subsubsection{Impact of Attribute Selection Phase}
Table~\ref{tab:attr_sel} reports results for a variant of LDI where tuples are randomly sampled, but only attributes in $(p,q)$-approximate dependency with the target attribute are retained. This setup isolates the effect of removing irrelevant attributes. Overall, by narrowing the input to the most informative data, the model can better focus on the patterns that truly matter. The results show that keeping only dependent attributes preserves, and in some cases improves, performance, confirming that our attribute selection phase successfully filters out distracting or redundant features. An especially notable improvement appears on the \textit{Zomato} dataset, which contains a larger number of attributes. For example, with $k=3$, exact match accuracy increases from 0.846 to 0.958 and ROUGE-1 F1 from 0.866 to 0.988 after removing irrelevant attributes. The gains are more evident when $k=3$, highlighting the importance of reducing input noise when only a few in-context examples are available. Beyond accuracy, this phase improves traceability and explainability. Limiting the input clarifies which attributes support each prediction. For instance, when imputing a missing city using only a phone number, any correct or incorrect prediction can be directly attributed to the LLM’s understanding of phone number patterns. This makes the model’s decisions easier to validate and debug.

\begin{table}[tb!]
\centering
\caption{Impact of attribute selection phase on accuracy with randomly selected tuples for $k=3$ (and $k=10$ in parentheses)}
\label{tab:attr_sel}
\smallfootnote

\newcommand{\pstack}[1]{%
  \shortstack{\rule{0pt}{2.0ex}#1\rule[-1.4ex]{0pt}{0pt}}%
}

\begin{tabular}{|
>{\columncolor{my_blue_color}}C{12mm}|
C{7.5mm}C{7.5mm}C{7.5mm}|
C{7.5mm}C{7.5mm}C{7.5mm}|}
\hline
\cellcolor{my_blue_color} 
& \multicolumn{3}{c|}{\cellcolor{my_blue_color}Exact Match} 
& \multicolumn{3}{c|}{\cellcolor{my_blue_color}ROUGE-1 F1} \\ \cline{2-7} 

\rowcolor{my_blue_color}
\multirow{-2}{*}{\cellcolor{my_blue_color}Dataset} 
& \pstack{All\\Attr.} 
& \pstack{Dep.\\Attr.} 
& Gain 
& \pstack{All\\Attr.} 
& \pstack{Dep.\\Attr.} 
& Gain \\ \hline

\cellcolor{my_blue_color}Buy 
& \pstack{0.923\\(0.928)} 
& \pstack{0.923\\(0.923)} 
& \pstack{0.0\%\\(-0.5\%)} 
& \pstack{0.985\\(0.985)} 
& \pstack{0.985\\(0.985)} 
& \pstack{0.0\%\\(0.0\%)} \\ \hline

\cellcolor{my_blue_color}Restaurant 
& \pstack{0.747\\(0.766)} 
& \pstack{0.761\\(0.772)} 
& \pstack{+1.9\%\\(+0.8\%)} 
& \pstack{0.894\\(0.890)} 
& \pstack{0.885\\(0.890)} 
& \pstack{-1.0\%\\(0.0\%)} \\ \hline

\cellcolor{my_blue_color}Zomato 
& \pstack{0.846\\(0.946)} 
& \pstack{0.958\\(0.964)} 
& \pstack{+13.2\%\\(+1.9\%)} 
& \pstack{0.866\\(0.966)} 
& \pstack{0.988\\(0.988)} 
& \pstack{+14.1\%\\(+2.3\%)} \\ \hline

\cellcolor{my_blue_color}Phone 
& \pstack{0.950\\(0.944)} 
& \pstack{0.952\\(0.942)} 
& \pstack{+0.2\%\\(-0.2\%)} 
& \pstack{0.980\\(0.978)} 
& \pstack{0.980\\(0.980)} 
& \pstack{0.0\%\\(+0.2\%)} \\ \hline

\end{tabular}
\end{table}

\subsubsection{Impact of Tuple Selection Phase}\label{sec:tup_sel}
Table~\ref{tab:tup_sel} compares two settings: random tuple selection and our proposed diverse similarity-based method. In our approach, similarity is computed explicitly over the attributes identified in phase one, ensuring that selected examples are similar in the most informative dimensions and avoiding irrelevant features that could dilute the closeness of the examples.
The results show that our similarity-based tuple selection leads to substantial improvements in exact match accuracy across all datasets and values of $k$. For example, in the \textit{Buy} dataset with $k=3$, accuracy increases from 0.923 (random) to 0.974 (diverse similarity). Similar trends are observed in the \textit{Phone} dataset (0.952 to 0.976) and \textit{Zomato} dataset (0.958 to 0.976). Likewise, in the more difficult \textit{Restaurant} dataset, the model achieves better performance, improving from 0.761 to 0.816 at $k=3$, and from 0.772 to 0.832 at $k=10$. Interestingly, while the accuracy improvements are often substantial, the changes in ROUGE scores are minimal with only minor differences across datasets.
This suggests that the similarity-based selection method helps the model make more precise and fully correct predictions, as opposed to simply improving surface-level similarity.
This precision stems from selecting relevant examples that reinforce correct specificity. For instance, when imputing brand names in the \textit{Buy} dataset, random examples may lack an LG product, causing the model to predict the generic “LG”. In contrast, our method is more likely to retrieve “LG Electronics”, yielding a more accurate output. Similar cases occur with “Sony” vs. “Sony Ericsson” in \textit{Phone}, “HSR” vs. “HSR Layout” in \textit{Zomato}, and “New York” vs. “New York City” in \textit{Restaurant} datasets.
Moreover, the tuple selection strategy implicitly leverages redundancy and repetition in the data. Many datasets contain repeated patterns, such as multiple tuples referring to the same location, brand, or product variant. By selecting a diverse yet similar set of examples, our method helps the LLM generalize more effectively from minimal context. This is particularly valuable in retrieval-augmented settings with limited in-context examples, where each example has a strong influence on the model’s output.
The ablation results confirm that both attribute and tuple selection are critical to the effectiveness of LDI. Phase one improves the model traceability and paves the way for effective tuple selection in the next phase. Phase two then selects diverse and similar examples that help the model better capture subtle variations in the data. Together, these phases guide the model with focused, representative context, leading to improved accuracy and more interpretable behavior~\footnote{We further analyze the zero-shot setting to compare against the LLM's internal knowledge; results are provided in Appendix~\ref{app:zeroshot}}.

\begin{table}[tb!]
\centering
\caption{Impact of the tuple selection phase on accuracy using only the dependent attributes for $k=3$ (and $k=10$ in parentheses)}
\label{tab:tup_sel}
\smallfootnote

\newcommand{\pstack}[1]{%
  \shortstack{\rule{0pt}{2.0ex}#1\rule[-1.4ex]{0pt}{0pt}}%
}

\begin{tabular}{|
>{\columncolor{my_blue_color}}C{12mm}|
C{8mm}C{8mm}C{8mm}|
C{8mm}C{8mm}C{8mm}|}
\hline
\cellcolor{my_blue_color}
& \multicolumn{3}{c|}{\cellcolor{my_blue_color}Exact Match}
& \multicolumn{3}{c|}{\cellcolor{my_blue_color}ROUGE-1 F1} \\ \cline{2-7}

\rowcolor{my_blue_color}
\cellcolor{my_blue_color}Dataset
& \pstack{Rand.}
& \pstack{Div.\\Sim.}
& Gain
& \pstack{Rand.}
& \pstack{Div.\\Sim.}
& Gain \\ \hline

\cellcolor{my_blue_color}Buy
& \pstack{0.923\\(0.923)}
& \pstack{0.974\\(0.974)}
& \pstack{+5.5\%\\(+5.5\%)}
& \pstack{0.985\\(0.985)}
& \pstack{0.991\\(0.991)}
& \pstack{+0.6\%\\(+0.6\%)} \\ \hline

\cellcolor{my_blue_color}Restaurant
& \pstack{0.761\\(0.772)}
& \pstack{0.816\\(0.832)}
& \pstack{+7.2\%\\(+7.8\%)}
& \pstack{0.885\\(0.890)}
& \pstack{0.887\\(0.889)}
& \pstack{+0.2\%\\(-0.1\%)} \\ \hline

\cellcolor{my_blue_color}Zomato
& \pstack{0.958\\(0.964)}
& \pstack{0.976\\(0.974)}
& \pstack{+1.9\%\\(+1.0\%)}
& \pstack{0.988\\(0.988)}
& \pstack{0.983\\(0.983)}
& \pstack{-0.5\%\\(-0.5\%)} \\ \hline

\cellcolor{my_blue_color}Phone
& \pstack{0.952\\(0.942)}
& \pstack{0.976\\(0.970)}
& \pstack{+2.5\%\\(+3.0\%)}
& \pstack{0.980\\(0.980)}
& \pstack{0.986\\(0.980)}
& \pstack{+0.6\%\\(0.0\%)} \\ \hline

\end{tabular}
\end{table}

\subsection{Varying p and q}\label{sec:varying-pq}
To assess the robustness of LDI under different noise levels in approximate dependencies, we varied $p$ and $q$ from 0 to 0.75 in steps of 0.25 and evaluated performance over five independent runs per setting. We used the \textit{Zomato} dataset, whose larger number of columns allows clearer observation of how attribute selection behaves under different noise levels, and we exclude settings where $p$ or $q$ exceed 0.75 because the dataset is highly noisy; stricter thresholds (e.g., $p, q =1$) would remove nearly all dependencies, often leaving no attributes selected. When either parameter is set to 0, no constraint is applied and all attributes are retained. Specifically, $q=0$ removes the frequency threshold for substring co-occurrence, and $p=0$ removes the requirement on the number of distinct target values supporting a dependency. 
Table~\ref{tab:pq_main} reports imputation performance when $k=3$ examples are selected, comparing (a) random selection and (b) diverse similarity-based selection. For both strategies, performance improves once $p,q \ge 0.25$, compared to the unfiltered baseline ($p=q=0$), and the improvement is more pronounced for random selection, where attribute filtering plays a stronger role.
With random selection (Table~\ref{tab:pq_main_rand}), the baseline setting ($p=q=0$) achieves 0.846 exact-match accuracy and 0.883 ROUGE-1, matching FMW results. Increasing both thresholds to 0.25 raises performance to roughly 0.960 accuracy and 0.978 ROUGE-1. Performance remains stable as $p$ and $q$ grow, even when the model uses only two attributes at $p=q=0.75$, reducing over 96\% of the input size while maintaining similar accuracy.
Using diverse similarity (Table~\ref{tab:pq_main_sim}) further improves performance across all parameter settings. Without attribute filtering ($p=q=0$), accuracy reaches 0.944 and ROUGE-1 0.951, demonstrating the benefit of selecting informative tuples alone. The best results occur at $p=0.75,q=0.50$ and $p=0.50,q=0.75$, achieving 0.978 accuracy and ROUGE-1 scores of 0.990 and 0.986. However, these configurations select more attributes than the stricter setting $p=q=0.75$, which uses only two attributes (see Table~\ref{tab:pq}) and achieves nearly identical performance. This highlights the flexibility of our method that tighter thresholds can greatly reduce input size with minimal impact on accuracy.

Table~\ref{tab:pq} further analyzes the effect of $p$ and $q$ on attribute reduction. When either is 0, all 16 attributes are selected, and as the thresholds increase, the selection becomes more restrictive, decreasing the average number of attributes from 7.6 at $p=q=0.50$ to 2 at $p=q=0.75$. This confirms that stricter dependency thresholds successfully remove weak and noisy attributes while preserving the key information needed for imputation.



\begin{table}[tb!]
\smallfootnote
\centering
\caption{Effect of varying $(p,q)$ on imputation performance in the Zomato dataset for $k=3$, with Exact Match accuracy (and ROUGE-1 score in parentheses)}
\begin{subtable}[t]{0.48\textwidth}
\centering
\caption{Example selection: Random}
\label{tab:pq_main_rand}
\begin{tabular}{|
>{\columncolor{my_blue_color}}C{7mm}|C{15mm}|C{14.5mm}|C{14.5mm}|C{14.5mm}|}
\hline
{\footnotesize q$\downarrow$~p$\rightarrow$} & \cellcolor{my_blue_color}0 & \cellcolor{my_blue_color}0.25 & \cellcolor{my_blue_color}0.50 & \cellcolor{my_blue_color}0.75 \\ \hline
0  & 0.846~(0.883)\textsuperscript{*} & 0.846~(0.883) & 0.846~(0.883) & 0.846~(0.883) \\ \hline
0.25 & 0.846~(0.883) & \cellcolor{green!50}0.960~(0.978) & \cellcolor{green!50}0.960~(0.977) & \cellcolor{green!50}0.954~(0.974) \\ \hline
0.50 & 0.846~(0.883) & \cellcolor{green!50}0.966~(0.982) & \cellcolor{green!50}0.960~(0.973) & \cellcolor{green!50}0.960~(0.975) \\ \hline
0.75 & 0.846~(0.883) & \cellcolor{green!50}0.954~(0.972) & \cellcolor{green!50}\textbf{0.970~(0.985)} & \cellcolor{green!50}0.960~(0.980) \\ \hline
\end{tabular}

\end{subtable}
\hfill
\begin{subtable}[t]{0.48\textwidth}
\centering
\bigskip
\caption{Example selection: Diverse similarity}
\label{tab:pq_main_sim}
\begin{tabular}{|
>{\columncolor{my_blue_color}}C{7mm}|C{15mm}|C{14.5mm}|C{14.5mm}|C{14.5mm}|}
\hline
{\footnotesize q$\downarrow$~p$\rightarrow$} & \cellcolor{my_blue_color}0 & \cellcolor{my_blue_color}0.25 & \cellcolor{my_blue_color}0.50 & \cellcolor{my_blue_color}0.75 \\ \hline
0  & 0.944~(0.951) & 0.944~(0.951) & 0.944~(0.951) & 0.944~(0.951) \\ \hline
0.25 & 0.944~(0.951) & \cellcolor{green!50}0.966~(0.979) & \cellcolor{green!50}0.968~(0.978) & \cellcolor{green!50}0.966~(0.977) \\ \hline
0.50 & 0.944~(0.951) & \cellcolor{green!50}0.964~(0.976) & \cellcolor{green!50}0.966~(0.978) & \cellcolor{green!50}\textbf{0.978~(0.990)} \\ \hline
0.75 & 0.944~(0.951) & \cellcolor{green!50}0.964~(0.976) & \cellcolor{green!50}\textbf{0.978~(0.986)} & \cellcolor{green!50}0.976~(0.983)\textsuperscript{\dag} \\ \hline
\end{tabular}

\end{subtable}

\vspace{0.5em}

\raggedright
\textsuperscript{ *} indicates the FMW result, and \dag\, indicates the LDI result. 

\label{tab:pq_main}
\end{table}

\begin{table}[]
\smallfootnote
\centering
\caption{Effect of varying $(p, q)$ on the average \# selected Attr. in the Zomato dataset}
\label{tab:pq}
\begin{tabular}{|
>{\columncolor{my_blue_color}}C{10mm}|C{8mm}|C{8mm}|C{8mm}|C{8mm}|}
\hline
q$\downarrow$ \;  {p$\rightarrow$} & \cellcolor{my_blue_color}0 & \cellcolor{my_blue_color}0.25 & \cellcolor{my_blue_color}0.50 & \cellcolor{my_blue_color}0.75 \\ \hline
0  & 16   & 16   & 16   & 16   \\ \hline
0.25 & 16   & \cellcolor{green!50}11.8 & \cellcolor{green!50}10.2 & \cellcolor{green!50}6.8  \\ \hline
0.50 & 16   & \cellcolor{green!50}11.4 & \cellcolor{green!50}7.6  & \cellcolor{green!50}3.4  \\ \hline
0.75 & 16   & \cellcolor{green!50}7.4  & \cellcolor{green!50}5    & \cellcolor{green!50}2    \\ \hline
\end{tabular}
\end{table}


\subsection{Limited-Resource Setting}

To assess the effectiveness of LDI in resource-constrained environments, we evaluate its performance using Llama 3.2 3B, a smaller, open-source language model with significantly fewer parameters than GPT-4o-mini. This experiment compares LDI with FMW, which uses all attributes and randomly selected examples for imputation, without any filtering.

As shown in Table~\ref{tab:llama}, LDI consistently outperforms the baseline across all datasets. These results show that LDI’s improvements are not dependent on large-scale language models; rather, its design effectively enhances performance through better input selection by providing the model with cleaner, more focused input. When smaller LLMs such as Llama 3.2 3B are used, focusing on localized input becomes especially important as these models have limited capacity to process and reason over lengthy inputs. If irrelevant attributes or uninformative examples are included, they can dilute the clues and distract the model from important patterns.

A particularly interesting case is the \textit{Zomato} dataset with $k=10$, where there is a large drop in performance for both LDI and FMW. This dataset contains many lengthy attributes, and including 10 examples results in a large input prompt. In such cases, the model struggles to extract the patterns. Interestingly, using only 3 examples yields much better performance, suggesting that more examples do not always lead to better results when the context is noisy or overloaded. Even in this challenging setting, LDI performs considerably better than the baseline (0.252 vs. 0.128 in accuracy, and 0.404 vs. 0.286 in ROUGE F1), demonstrating its advantage in isolating informative content. These findings illustrate that guiding the model with concise, high-quality input, rather than simply more input, is key to achieving high performance.

\begin{table}[tb]
\centering
\caption{Performance comparison when deployed with a smaller language model (Llama 3.2 3B) for $k=3$ (and $k=10$ in parentheses)}
\label{tab:llama}
\smallfootnote

\newcommand{\pstack}[1]{%
  \shortstack{\rule{0pt}{2.0ex}#1\rule[-1.4ex]{0pt}{0pt}}%
}

\begin{tabular}{|
>{\columncolor{my_blue_color}}C{12mm}|
C{8mm}C{8mm}C{8mm}|
C{8mm}C{8mm}C{8mm}|}
\hline
\cellcolor{my_blue_color} 
& \multicolumn{3}{c|}{\cellcolor{my_blue_color}Exact Match} 
& \multicolumn{3}{c|}{\cellcolor{my_blue_color}ROUGE-1 F1} \\ \cline{2-7}

\rowcolor{my_blue_color}
\cellcolor{my_blue_color}Dataset 
& \cellcolor{my_blue_color}FMW 
& \cellcolor{my_blue_color}LDI 
& Gain 
& \cellcolor{my_blue_color}FMW 
& \cellcolor{my_blue_color}LDI 
& Gain \\ \hline

\cellcolor{my_blue_color}Buy
& \pstack{0.918 \\ (0.923)}
& \pstack{0.954 \\ (0.944)}
& \pstack{+3.9\% \\ (+2.3\%)}
& \pstack{0.954 \\ (0.959)}
& \pstack{0.980 \\ (0.973)}
& \pstack{+2.7\% \\ (+1.5\%)} \\ \hline

\cellcolor{my_blue_color}Restaurant
& \pstack{0.701 \\ (0.738)}
& \pstack{0.775 \\ (0.770)}
& \pstack{+10.6\% \\ (+4.3\%)}
& \pstack{0.802 \\ (0.833)}
& \pstack{0.837 \\ (0.839)}
& \pstack{+4.4\% \\ (+0.7\%)} \\ \hline

\cellcolor{my_blue_color}Zomato
& \pstack{0.766 \\ (0.128)}
& \pstack{0.894 \\ (0.252)}
& \pstack{+16.7\% \\ (+96.9\%)}
& \pstack{0.826 \\ (0.286)}
& \pstack{0.928 \\ (0.404)}
& \pstack{+12.3\% \\ (+41.3\%)} \\ \hline

\cellcolor{my_blue_color}Phone
& \pstack{0.910 \\ (0.854)}
& \pstack{0.966 \\ (0.964)}
& \pstack{+6.2\% \\ (+12.9\%)}
& \pstack{0.941 \\ (0.890)}
& \pstack{0.977 \\ (0.977)}
& \pstack{+3.8\% \\ (+9.8\%)} \\ \hline

\end{tabular}
\end{table}

\section{Conclusion and Future Work}\label{sec:con}
We proposed a localized imputation method that improves both accuracy and explainability of LLM-based data imputation by selecting informative attributes and representative tuples for each missing value. This focused approach reduces noise, handles inconsistencies, 
and guides the LLM toward more accurate predictions. Experiments on four real-world datasets show state-of-the-art performance with better explainability than existing methods.
Future work includes developing more advanced selection strategies where dependencies are conditional or involve more complex patterns, exploring transfer learning across domains, and providing richer explanations of model reasoning.

\begin{acks}
We sincerely thank the reviewers for their comments and suggestions, which improved the quality of this work.
This research was funded by the Natural Sciences and Engineering Research Council of Canada (NSERC).
Soroush Omidvartehrani was supported by the Alberta Innovates Graduate Student Scholarship (AIGSS).
\end{acks}


\bibliographystyle{ACM-Reference-Format}
\bibliography{r}

\clearpage
\appendix

\section{Detailed Performance Analysis}\label{app:analysis}

This appendix provides the detailed time complexity analysis of the main components
of LDI, including supporting lemmas and step-by-step derivations.

\subsection{Supporting Lemmas}

We first present two lemmas that are used in the analysis of both the attribute selection and tuple selection phases.

\begin{lemma}[Frequent Substring Detection]\label{lem:gst-app}
Given a collection of $n$ strings, each of length $\ell$, all substrings that appear in at least $q$ fraction of the strings can be found in time $\mathcal{O}(n \cdot \ell + r)$ and space $\mathcal{O}(n \cdot \ell)$, where $r$ is the number of substrings reported.
\end{lemma}

\begin{proof}
We build a Generalized Suffix Tree (GST) for the $n$ input strings, each of length $\ell$, after appending a unique terminal symbol to each string. This tree can be constructed in $\mathcal{O}(n \cdot \ell)$ time and space using Ukkonen’s algorithm~\cite{ukkonen1995line}. Each leaf is labeled with its corresponding string ID, and a post-order traversal is used to annotate each internal node with the set of string IDs in its subtree. For each node, if the number of distinct string IDs is at least $\lceil qn \rceil$, we report the substring represented by that node. Reporting all such substrings takes $\mathcal{O}(r)$ time, where $r$ is the number of substrings found. Therefore, the total time and space complexity is $\mathcal{O}(n \cdot \ell + r)$ and $\mathcal{O}(n \cdot \ell)$, respectively.
\end{proof}

\begin{lemma}[Uniqueness Across Groups]\label{lem:unique-app}
Given a collection of $s$ strings, each of length $\ell$, partitioned into $m$ groups, the time to determine whether each string appears uniquely within a single group is $\mathcal{O}(s \cdot \ell)$.
\end{lemma}

\begin{proof}
To determine whether each string appears in only one group, we use a hashmap that associates each string with the set of group identifiers in which it appears. We iterate over all $s$ strings and, for each one, insert it into the hashmap and record its group identifier. Since each string has length $\ell$, inserting or comparing a string in the hashmap takes $\mathcal{O}(\ell)$ time. Processing all $s$ strings therefore takes $\mathcal{O}(s \cdot \ell)$ time. After building the hashmap, we scan its entries to check whether any string is associated with more than one group. As there are at most $s$ unique strings, this step takes $\mathcal{O}(s)$ time. Hence, the overall time complexity is $\mathcal{O}(s \cdot \ell)$.
\end{proof}

\subsection{Attribute Selection}\label{app:attr}

The attribute selection phase depends on the number of candidate attributes $a_c$, the number of groups $m$ induced by the target attribute, the number of sampled tuples per group $n$, the average string length $\ell$, and the number of extracted substrings $r$ per group.

\paragraph{Step 1: Group Sampling.}
Tuples are grouped by their values in the target attribute and a subset of $m$ groups is sampled, with $n$ tuples selected per group. This requires a single pass over the dataset and takes $\mathcal{O}(|\mathcal{D}|)$ time.

\paragraph{Step 2: Pattern Detection.}
For each candidate attribute and each group, we detect substrings that appear in at least a $q$ fraction of the sampled tuples using a generalized suffix tree. By Lemma~\ref{lem:gst-app}, this step takes $\mathcal{O}(n \cdot \ell + r)$ time per group.
Across all $m$ groups and $a_c$ candidate attributes, the total cost is
\[
\mathcal{O}(a_c \cdot m \cdot (n \cdot \ell + r)).
\]

\paragraph{Step 3: Uniqueness Checking.}
All substrings extracted from different groups are checked for uniqueness across groups. Letting $s = m \cdot r$ be the total number of substrings per attribute, Lemma~\ref{lem:unique-app} implies a cost of $\mathcal{O}(s \cdot \ell)$ per attribute.
Across all candidate attributes, this results in a total time of
\[
\mathcal{O}(a_c \cdot m \cdot r \cdot \ell).
\]

\paragraph{Step 4: Optional Redundancy Filtering.}
Optionally, substrings that are fully contained within longer ones are removed within each group to improve interpretability. This requires pairwise comparisons among the $r$ substrings in a group, yielding a cost of $\mathcal{O}(r^2 \cdot \ell)$ per group.
Across all groups and candidate attributes, the total cost is
\[
\mathcal{O}(a_c \cdot m \cdot r^2 \cdot \ell).
\]

\paragraph{Overall Complexity.}
Combining all steps, the total time complexity of the attribute selection phase is:
\[
\mathcal{O}\!\left(|\mathcal{D}| + a_c \cdot m \cdot \ell \cdot (n + r)\right)
\]
when redundancy filtering is disabled, and
\[
\mathcal{O}\!\left(|\mathcal{D}| + a_c \cdot m \cdot \ell \cdot (n + r^2)\right)
\]
when redundancy filtering is enabled.

\subsection{Tuple Selection}\label{app:tuple}

To impute a missing value in a tuple $t_i$, LDI selects $k$ similar and diverse complete tuples. Let $c$ be the number of complete tuples (i.e., tuples with known values in $A_T$) and $a_s$ the number of selected attributes.

For each selected attribute, we build a generalized suffix tree over the attribute values of $t_i$ and all $c$ complete tuples. By Lemma~\ref{lem:gst-app}, constructing the GST takes $\mathcal{O}(c \cdot \ell)$ time per attribute. After the tree is built, the longest common substring between $t_i$ and any complete tuple can be computed in $\mathcal{O}(\ell)$ time.

Thus, computing similarity scores between $t_i$ and all $c$ complete tuples across $a_s$ attributes takes $\mathcal{O}(a_s \cdot c \cdot \ell)$ time. The $c$ tuples are then sorted by similarity in $\mathcal{O}(c \log c)$ time, and a final scan ensures diversity among the top-$k$ selected tuples. Therefore, the total time complexity of tuple selection for one incomplete tuple is
\[
\mathcal{O}(a_s \cdot c \cdot \ell + c \log c).
\]


\section{Implementation and Experiment Details}\label{app:exp}
\subsection{Datasets and Evaluation Setup}\label{app:exp-setup}

We evaluate LDI on four real-world datasets from three domains: Buy, Restaurant, Zomato, and Phone. The incomplete attributes are categorical, with vocabulary sizes ranging from 49 to 385. Buy and Restaurant have fewer attributes but are standard benchmarks~\cite{narayan2022can,chen2024seed,song2020imputing,mei2021capturing}, while Zomato and Phone have more attributes, enabling a comprehensive evaluation of attribute selection and scalability. Experiments for large datasets were limited to 1,000 rows to manage LLM calls.

Experiments were conducted on Ubuntu 20.04 with an AMD Ryzen 9 3900X (12-core) and 64 GB RAM, using Python 3.9.20. Group sampling was configured with $m=10$ groups and $n=10$ samples per group, with the maximum available used for smaller datasets. Inner threshold $q$ and outer threshold $p$ were set to 0.9 for Buy and Phone, and 0.8 for Zomato and Restaurant, allowing minor noise while maintaining strong evidence for dependencies~\footnote{Section~\ref{sec:varying-pq} reports a sensitivity analysis showing that LDI remains robust across a wide range of $p$ and $q$ values.}. 
We report results with $k=3$ and $k=10$ in-context examples, matching the default settings of FMW for fair comparison, and baseline hyperparameters follow the ranges reported in the original papers.
LLM-based imputation primarily used GPT-4o-mini (version 2024-07-18), with Llama 3.2 3B included to evaluate performance under constrained resources. GPT-4o-mini provides higher accuracy, while Llama 3.2 3B shows LDI’s portability and generalizability.

\subsection{LLM Prompt Template} \label{app:prompt}

Each input is serialized as key--value pairs and organized into three parts:
a task description, example tuples, and a query tuple with a missing target value.

\begin{tcolorbox}[colframe={my_blue_color},colback={my_blue_color},
left=3pt, right=3pt, top=3pt, bottom=3pt]
\small
\begin{verbatim}
[Context]
  A brief description of the imputation task.
[Examples]
  Example 1:
  Attr. A: Val. A1, Attr. B: Val. B1, ...
  Target Attr.: Target Val. 1
  Example 2:
  Attr. A: Val. A2, Attr. B: Val. B2, ...
  Target Attr.: Target Val. 2
  ...
[Query]
  Attr. A: Val. Ax, Attr. B: Val. Bx, ...
  Target Attr.: ?
\end{verbatim}
\end{tcolorbox}

\subsection{Zero-shot Baseline} \label{app:zeroshot}

To better understand the contribution of LDI beyond the inherent knowledge of LLMs, we evaluate a zero-shot baseline in which the model is queried without any retrieved examples (i.e., $k=0$) or attribute selection. In this setting, the LLM relies solely on its pretrained knowledge to infer missing values.

\begin{table}[bt!]
\smallfootnote
\centering
\caption{Zero-shot vs. LDI (k=10)}
\label{tab:zeroshot}
\begin{tabular}{|
>{\columncolor{my_blue_color}}C{20mm}|
C{20mm}|C{20mm}|}
\hline
Dataset & \cellcolor{my_blue_color}Zero-shot & \cellcolor{my_blue_color}LDI \\ \hline
Buy & 0.909 (0.951) & 0.974 (0.991) \\ \hline
Restaurant & 0.794 (0.858) & 0.832 (0.889) \\ \hline
Zomato & 0.146 (0.657) & 0.974 (0.983) \\ \hline
Phone & 0.945 (0.961) & 0.970 (0.980) \\ \hline
\end{tabular}
\end{table}

Table~\ref{tab:zeroshot} compares the exact match accuracy (and ROUGE-1 score in parentheses) of this baseline with LDI using $k=10$ examples. While zero-shot prompting achieves competitive performance on some datasets, its behavior is inconsistent and often lacks precision. In particular, without access to relevant examples, the model tends to generate plausible but less specific predictions, as it cannot ground its output in dataset-specific patterns or variations.
This effect is consistent with the observations in Section~\ref{sec:tup_sel}: providing relevant and similar examples enables the model to produce more precise and fully correct predictions, rather than approximate or generic ones. By selecting examples that reflect the underlying data distribution, LDI reinforces correct specificity and reduces ambiguity in the generated outputs.
This difference is especially pronounced in the Zomato dataset, where zero-shot performance is substantially lower. In such cases, accurate imputation depends on subtle, dataset-specific signals that are not captured by general knowledge alone. By contrast, LDI retrieves informative examples that expose these patterns, enabling the model to make accurate and contextually grounded predictions.
Overall, these results show that the gains of LDI stem not only from leveraging LLM capabilities, but from guiding the model with relevant, localized examples that improve precision and reliability.





\end{document}